\input harvmac
\input amssym

\lref\gaugemediation{ G.~F.~Giudice and R.~Rattazzi,
  ``Theories with gauge-mediated supersymmetry breaking,''
  Phys.\ Rept.\  {\bf 322}, 419 (1999)
  [hep-ph/9801271].
}

\lref\murayamaetal{
  G.~F.~Giudice, M.~A.~Luty, H.~Murayama and R.~Rattazzi,
  ``Gaugino mass without singlets,''
  JHEP {\bf 9812}, 027 (1998)
  [arXiv:hep-ph/9810442].
}

\lref\randallsundrum{
  L.~Randall and R.~Sundrum,
  ``Out of this world supersymmetry breaking,''
  Nucl.\ Phys.\ B {\bf 557}, 79 (1999)
  [arXiv:hep-th/9810155].
}

\lref\baggeretal{
  J.~A.~Bagger, T.~Moroi and E.~Poppitz,
  ``Anomaly mediation in supergravity theories,''
  JHEP {\bf 0004}, 009 (2000)
  [arXiv:hep-th/9911029].
  }

\lref\AffleckXZ{
  I.~Affleck, M.~Dine and N.~Seiberg,
  ``Dynamical Supersymmetry Breaking In Four-Dimensions And Its
  Phenomenological Implications,''
  Nucl.\ Phys.\ B {\bf 256}, 557 (1985).
}

\lref\NelsonNF{
  A.~E.~Nelson and N.~Seiberg,
  ``R symmetry breaking versus supersymmetry breaking,''
  Nucl.\ Phys.\ B {\bf 416}, 46 (1994)
  [arXiv:hep-ph/9309299].
}

\lref\baggeretal{
  J.~A.~Bagger, T.~Moroi and E.~Poppitz,
  ``Anomaly mediation in supergravity theories,''
  JHEP {\bf 0004}, 009 (2000)
  [arXiv:hep-th/9911029].
  J.~A.~Bagger, T.~Moroi and E.~Poppitz,
  ``Quantum inconsistency of Einstein supergravity,''
  Nucl.\ Phys.\ B {\bf 594}, 354 (2001)
  [arXiv:hep-th/0003282].
}

  \lref\macintire{ 
  M.~Dine and D.~MacIntire,
  ``Supersymmetry, Naturalness, And Dynamical Supersymmetry Breaking,''
  Phys.\ Rev.\ D {\bf 46}, 2594 (1992)
  [arXiv:hep-ph/9205227].
  }

\lref\cremmer{ 
  E.~Cremmer, S.~Ferrara, L.~Girardello and A.~Van Proeyen,
   ``Yang-Mills Theories With Local Supersymmetry: Lagrangian, Transformation
  Laws And Superhiggs Effect,''
  Nucl.\ Phys.\ B {\bf 212}, 413 (1983).}

\lref\BoydaNH{
  E.~Boyda, H.~Murayama and A.~Pierce,
  ``DREDed anomaly mediation,''
  Phys.\ Rev.\ D {\bf 65}, 085028 (2002)
  [arXiv:hep-ph/0107255].
}

\lref\wessbagger{J. Wess and J. Bagger, {\it Supersymmetry and
supergravity}, (1992) Princeton Univ. Press, Princeton, N.J. }

\lref\grisaru{
  M.~T.~Grisaru, W.~Siegel and M.~Rocek,
  ``Improved Methods For Supergraphs,''
  Nucl.\ Phys.\ B {\bf 159}, 429 (1979).
  }

\lref\IntriligatorDD{
  K.~Intriligator, N.~Seiberg and D.~Shih,
  ``Dynamical SUSY breaking in meta-stable vacua,''
  JHEP {\bf 0604}, 021 (2006)
  [arXiv:hep-th/0602239].
}

\lref\vy{
  G.~Veneziano and S.~Yankielowicz,
  ``An Effective Lagrangian For The Pure N=1 Supersymmetric Yang-Mills
  Theory,''
  Phys.\ Lett.\ B {\bf 113}, 231 (1982).
}

 \lref\martinvaughn{
  S.~P.~Martin and M.~T.~Vaughn,
   ``Two Loop Renormalization Group Equations For Soft Supersymmetry Breaking
  Couplings,''
  Phys.\ Rev.\ D {\bf 50}, 2282 (1994)
  [arXiv:hep-ph/9311340].
}

 \lref\split{ 
  N.~Arkani-Hamed and S.~Dimopoulos,
  ``Supersymmetric unification without low energy supersymmetry and  signatures
  for fine-tuning at the LHC,''
  JHEP {\bf 0506}, 073 (2005)
  [arXiv:hep-th/0405159].
  N.~Arkani-Hamed, S.~Dimopoulos, G.~F.~Giudice and A.~Romanino,
  Nucl.\ Phys.\ B {\bf 709}, 3 (2005)
  [arXiv:hep-ph/0409232].
 }

\lref\adsreview{
  O.~Aharony, S.~S.~Gubser, J.~M.~Maldacena, H.~Ooguri and Y.~Oz,
  ``Large N field theories, string theory and gravity,''
  Phys.\ Rept.\  {\bf 323}, 183 (2000)
  [arXiv:hep-th/9905111].
}

\lref\weinberg{S. Weinberg, {\it The Quantum Theory of Fields}, Cambridge University Press (1995),
Cambridge.}

\lref\sv{
  M.~A.~Shifman, A.~I.~Vainshtein and V.~I.~Zakharov,
  Phys.\ Lett.\ B {\bf 166}, 334 (1986).
}

\def\dslash{\not{\hbox{\kern-2pt $\partial$}}}
\def\Dslash{\not{\hbox{\kern-4pt $D$}}}
\def\Oslash{\not{\hbox{\kern-4pt $O$}}}
\def\Qslash{\not{\hbox{\kern-4pt $Q$}}}
\def\pslash{\not{\hbox{\kern-2.3pt $p$}}}
\def\kslash{\not{\hbox{\kern-2.3pt $k$}}}
\def\lslash{\not{\hbox{\kern-2.3pt $l$}}}
\def\qslash{\not{\hbox{\kern-2.3pt $q$}}}
\def\epsilonslash{\not{\hbox{\kern-2.3pt $\epsilon$}}}

\def\m3{m_{3/2}}

\def\K3{{\bf K3}}
\def\journal#1&#2(#3){\unskip, \sl #1\ \bf #2 \rm(19#3) }
\def\andjournal#1&#2(#3){\sl #1~\bf #2 \rm (19#3) }

\def\bar{\overline}

\def\frac#1#2{{#1\over#2}}

\def\inbar{\,\vrule height1.5ex width.4pt depth0pt}
\def\IC{\relax\hbox{$\inbar\kern-.3em{\rm C}$}}
\def\IR{\relax{\rm I\kern-.18em R}}
\def\IP{\relax{\rm I\kern-.18em P}}

%
%

%
\catcode`\@=11
\def\slash#1{\mathord{\mathpalette\c@ncel{#1}}}
\overfullrule=0pt

\def\underrel#1\over#2{\mathrel{\mathop{\kern\z@#1}\limits_{#2}}}

\catcode`\@=12


%

\def\det{{\rm det}}

\def\det{{\rm det}}



%
%
%
%
%
%

\Title{\vbox{\baselineskip12pt\hbox{}
\hbox{SCIPP-2006/14}\hbox{UCI-TR-2006-14 }} } {\vbox{
\centerline{Comments on Quantum Effects in }
\centerline{Supergravity Theories} }}
\bigskip
\bigskip
\centerline{Michael Dine$^1$ and Nathan Seiberg$^2$}
\bigskip
\centerline{$^1$ {\it Santa Cruz Institute for Particle Physics,
Santa Cruz CA 95064, USA}} \centerline{$^2${\it School of Natural
Sciences, Institute for Advanced Study, Princeton, New Jersey
08540} }
\bigskip
\bigskip
\noindent {\centerline {\it Abstract}}
 We elucidate the physics
underlying ``anomaly mediation'', giving several alternative
derivations of the formulas for gaugino and scalar masses.  We
stress that this phenomenon is of a type familiar in field theory,
and does not represent an anomaly, nor does it depend on
supersymmetry breaking and its mediation. Analogous phenomena are
common in QFT and this particular phenomenon occurs also in
supersymmetric theories without gravity.

\bigskip
\Date{January 2007}


\newsec{Introduction}

The soft breaking terms in the MSSM depend on the supersymmetry
breaking mechanism and on the mediation of that breaking to the
fields in the MSSM.  One of the most popular mediation schemes is
a special case of gravity mediation which is generally referred to
as ``anomaly mediation" \refs{\murayamaetal,\randallsundrum}. The
anomaly mediation contribution to gaugino masses in theories with
dynamical supersymmetry breaking is often the largest
contribution. If the leading K\"ahler potential has a particular,
{\it sequestered}, form \randallsundrum, then the anomaly
mediation contribution to scalar masses is the largest one and it
leads to universal (if problematic) masses.

But there is much that is puzzling about these contributions:
 \item{1.} The {\it anomaly} at issue is an anomaly in conformal
 transformations, which are not symmetries of the theory. What
 does it mean to have an anomaly in a symmetry which is not
 present classically?  We will argue that the underlying
 phenomenon can be understood without relying on the conformal
 anomaly; in fact, we will show that the relevant phenomenon is
 not that of an anomaly.
 \item{2.} All discussions of the problem are set within
 supergravity models, and most are tied to a very particular
 formulation of supergravity theories which uses conformal
 compensators, hence the relation to the conformal anomaly.
 We will work in components without using any particular set of
 auxiliary fields.  We will see that {\it the phenomenon
 already arises in globally supersymmetric theories}.  It has no
 fundamental connection to local supersymmetry, much less any
 particular supergravity formalism.
 \item{3.} It is unclear, in the usual presentations, whether
 the effect should be understood in a Wilsonian effective
 action with a cutoff at a given energy $\Lambda$, or in a
 1PI effective action.  Related to this is the question whether
 the phenomenon is an ultraviolet (UV) or an infrared (IR)
 effect.  We will show that the issue is a {\it local
 counterterm} which is needed in order to preserve supersymmetry.
 It is associated with physics {\it near the cutoff $\Lambda$
 of the Wilsonian action}, and should be thought of as an
 ultraviolet effect.

This note seeks to clarify the nature of these phenomena. We will
explain our assertion that the effect is not an anomaly, nor is it
intrinsically gravitational.  We will demonstrate
that it arises in theories where
supersymmetry is unbroken, as well as in theories of supersymmetry
breaking.  The distinctive feature is the appearance of a contact
term coupling the superpotential to gaugino bilinears. This term
is not supersymmetric invariant.  But this lack of invariance is
cancelled by a the lack of invariance of the measure of the light
fields. As we explain, this sort of phenomenon is already familiar
in ordinary QED in two and four dimensions. We will, as a result,
adopt a different terminology, referring to these phenomena as the
{\it gaugino counterterm}.

Some experts might feel that our explanations are not new and we
merely review known facts.  However, a careful examination of the
published literature and numerous detailed discussions with many
physicists convinced us that the subject is confusing and deserves
a new, clear exposition.

There are two important issues which we will not address. The
first, is the question of whether the sequestered form of the
K\"ahler potential is natural or not.  The second, is whether the
sequestered form leads to an acceptable phenomenology.
Instead, we will focus only on the more formal field
theoretic issues associated with the counterterms.

The rest of this paper is organized as follows. In the next
section, we discuss at some length the {\it gaugino counterterm},
the crucial element of anomaly mediation. We explain that this
counterterm arises from integrating out modes near the UV cutoff
of the Wilsonian effective action.  It can be thought of either as
a contribution of the regulator fields at that scale or as a local
term which is explicitly present in that action.  In the latter
case it seems to violate supersymmetry, but this symmetry is
restored through the interactions of the light fields. We remind
the reader that this phenomenon is familiar in QED both in two and
four dimensions.  In section three we give two derivations of the
counterterms.  One is a review of a standard derivation based on
Pauli-Villars fields. The second considers theories in which the
gauge symmetry is spontaneously broken, and the low energy
effective action is completely local. In this setup, requirement
of the local counterterm is almost obvious.  In section four, we
show that the gaugino counterterm already arises in globally
supersymmetric theories. In section 5, we introduce simple models
of supersymmetry breaking which incorporate features of models
with dynamical supersymmetry breaking.   Section six is a
discussion of the scalar counterterms from this point of view.  In
an appendix, we discuss two and four dimensional electrodynamics
in a manner which stresses the parallels to the gaugino
counterterm.

\newsec{The Gaugino Counter Term}

The gaugino counterterm, the phenomenon which underlies anomaly
mediation, can be understood in a variety of ways.  Historically,
there have been several derivations. In \macintire, it was
observed that, at least naively, in a theory with broken
supersymmetry, the presence of heavy fields leads to one loop
diagrams contributing to gaugino masses, even when the gauge
coupling function is trivial, and no such term is permitted by
local supersymmetry in the effective action.  Regulating the
diagrams with a Pauli-Villars field eliminates these terms, but
this raises the question:  why there are no such contributions
from {\it massless} fields?  In
\refs{\murayamaetal,\randallsundrum} it was shown that such
contributions are in fact present.  These derivations of gaugino
masses were tied to a particular supergravity formalism, and to
anomalous field redefinitions in that formalism. The authors of
\randallsundrum\ exhibited additional contributions at two loops
to scalar masses as well, beyond those expected from the local
supergravity action. Reference \baggeretal\ offered an explanation
of this phenomenon in terms of anomalies in various field
redefinitions in supergravity.  They summarized their analysis in
terms of non-local operators in the 1PI effective action.

These derivations are all correct, but in each case their physical
significance is obscure.  Our goal is to find a more satisfying
conceptual setting. For this purpose, it is enough to consider
supersymmetric QED; i.e.\ a $U(1)$ gauge theory with two chiral
superfields $\phi_\pm$ with opposite $U(1)$ charges. We will
include a nonzero constant $W_0$ in the superpotential.  Because
of its simplicity, we first review in this section the
Pauli-Villars analysis, showing that a contact term proportional
to $W^* \lambda \lambda$ is generated.  Such a term arises whether
or not supersymmetry is broken. {\it This gaugino contact term
cannot arise, however, as a term in a supersymmetric effective
action}.  This is the would-be paradox. We explain in this
section, and in the appendix, that this term is of a type quite
familiar in ordinary field theory, even in QED. It does not signal
the presence of an anomaly.  In the next section, we briefly
review the derivation using the superconformal compensator, and
then provide a more transparent derivation, in a theory free of
non-trivial infrared physics.

Turning to the U(1) model, if supersymmetry is unbroken the
cosmological constant is negative and the ground state is AdS; if
supersymmetry is broken (due to some additional, hidden sector
fields, say) we can use $W_0$ to set the cosmological constant to
zero.  The superpotential is
 \eqn\uonew{ W= W_0 + m \phi_+ \phi_-}
and we will take, for simplicity, the constants
 \eqn\recons{W_0=\m3 M_p^2 \qquad {\rm and} \qquad m}
to be real and $m \gg \m3$, such that the AdS radius $R_{AdS}={1
\over \m3}$ is large compared to the inverse mass, and the space
is approximately flat.  The K\"ahler potential is simply
 \eqn\uonek{ K = \bar \phi_{+}\phi_+ + \bar \phi_{-}\phi_-.}

The supergravity potential is
 \eqn\AdSpot{\eqalign{
 V = & \ e^{K/M_p^2} \left[ g^{i\bar j}\left(\partial_i W +
 {1\over M_p^2} \partial_i K~ W\right)\left(\partial_{\bar i}
 \bar W + {1\over M_p^2} \partial_{\bar i} K ~\bar W\right)
 - 3 {1\over M_p^2} W\bar W\right]\cr
 = & -3 \m3^2 M_p^2 + (m^2 - 2 \m3^2)(|\phi_+|^2 + |\phi_-|^2)
 - \m3 m(\phi_+\phi_- + \bar\phi_+\bar\phi_-) + ...}}
where we neglected terms which include higher powers of the
fields. The first term is the cosmological constant and the other
terms lead to scalar masses and interactions. Note in particular
that the interaction with nonzero $W_0$ leads both to
contributions to scalar masses of the form $|\phi_\pm|^2$ and to
B-terms of the form $\phi_+\phi_- + c.c.$. The mass eigenstates
are $\phi_{1,2}={1\over \sqrt 2}(\phi_+ \mp \bar \phi_-)$ with
eigenvalues
 \eqn\AdSmasses{ m_{1,2}^2=m^2 -2 \m3^2 \pm m \m3.}
A simple calculation shows that the masses of the fermionic
partners of $\phi_\pm$ are not modified by the interaction with
$W_0$ and are simply $m$. We see that the masses of the two bosons
and the fermion are not degenerate even though supersymmetry is
not broken\foot{ This agrees with the expression for the
corresponding dimensions in the three dimensional boundary
conformal field theory.  The dimensions of the fermion operator
and the boson operators are \adsreview:
 \eqn\sclardim{\eqalign{
 &\Delta_F= {3\over 2} + {\vert m_F \vert \over \m3} = {3\over 2} +
 {\vert m\vert \over \m3} \cr
 &\Delta_{B1,2} = {3\over 2} + \sqrt{{9\over 4} +
 {m_{1,2}^2 \over \m3^2}}= {3\over 2} + {\vert m \vert \over \m3} \pm
 {1\over  2} = \Delta_F \pm {1\over 2} .}}
The dimensions $\Delta_F$ and $\Delta_{B\pm} = \Delta_F \pm
{1\over 2}$ are such that the corresponding operators are in the
same supersymmetry multiplet.}.

With these masses, there is a one loop graph for the photino mass
$m_\lambda$.  Using the B-term in \AdSpot\ we find
 \eqn\gamass{m_\lambda = e^2 \int {d^4p \over (2 \pi)^4} {\m3 m^2
 \over (p^2 + m^2)^3} = {e^2 \over 16 \pi^2} \m3}
and therefore there is an effective interaction
 \eqn\uonecontact{{\alpha \over 4\pi M_p^2} W_0^*\lambda \lambda}
($\lambda$ is the photino).  Note that the expression \gamass\ is
finite, does not need regularization and seems unambiguous.

This result appears paradoxical, since no such term appears in the
general supergravity action \refs{\cremmer,\wessbagger}.  How can
loops generate terms which do not respect the symmetry of the
theory, and therefore cannot be present in an invariant
Lagrangian? The answer is that this term can be cancelled by a
local counterterm. This is easily seen if we regulate the theory
using a supersymmetric regulator like Pauli-Villars with mass
$\Lambda$. The contribution of \uonecontact\ is independent of the
mass $m$, and similarly the contribution of the regulator is
finite and is independent of $\Lambda$.  Since the Pauli-Villars
field contributes with the opposite sign, its contribution exactly
cancels that of the fields $\phi_\pm$ \uonecontact, and we end up
without that term!

Let us consider now the massless theory with $m=0$.  Here \gamass\
vanishes, but the regulator contribution is nonzero and we are
left with a term of the form of \uonecontact, but with the
opposite sign. This is the famed ``anomaly mediated" gaugino mass,
in the special case of a $U(1)$ theory (and with {\it unbroken}
supersymmetry).

To summarize, in the massive theory a term like \uonecontact\ is
not present, but it is present in the massless theory.

What is the Wilsonian action interpretation of this result?
Consider first the massive theory with a UV cutoff $\Lambda \gg
m$.  The matter fields are much lighter than the UV cutoff and
their loops are not included in the effective action.  If we
regularize the theory with Pauli-Villars fields with mass
$\Lambda$, these fields lead to a gaugino counterterms. Therefore,
this term is generated by physics near the UV cutoff $\Lambda$,
but is not explicitly present in the Wilsonian action. If
alternatively, we use a sharp momentum cutoff, then this term must
be introduced ``by hand."  The sharp momentum cutoff is not
supersymmetric, and therefore it is not surprising that such a
counterterm is needed\foot{The case of supersymmetric dimensional
reduction was analyzed in \murayamaetal\ and in more detail in
\BoydaNH.  There it was argued that to respect the local conformal
invariance of the supergravity construction with $4-\epsilon$
spatial coordinates,  certain operators with coefficients of order
$\epsilon$ have to be added to the action. These operators lead to
the counterterm at one loop.  One can view these operators as
added counterterms required for supersymmetry. Alternatively, we
can leave out these operators and add the gaugino counterterm ``by
hand."}. One way of thinking about this situation is that the
regulated measure of the light fields is not supersymmetric and
supersymmetry is restored only by adding this local counterterm.

Regardless of whether we study the system with a supersymmetric
regulator, where the term arises from the physics around
$\Lambda$, or with a nonsupersymmetric cutoff, where it is a local
counterterm, we see here cancellations between contributions
arising at different momentum scales.

It should be stressed that this phenomenon is not an anomaly.  An
{\it anomaly} is a lack of symmetry, which {\it cannot be restored
with local counterterms}.  Here, with some cutoffs the effective
lagrangian is supersymmetric; with some it is not, but the
symmetry can be restored by adding a local counterterm.

As the cutoff is further reduced to be of order $m$ we should take
into account the matter loops which cancel this contribution.
Finally, for very low cutoff $\Lambda \ll m$ the theory includes
only the photon multiplet and the counterterm is not present.  We
conclude that this gaugino counterterm is present only at energies
above $m$, but it is absent at energies below $m$.

Since the counterterm is present at energies larger than $m$, it
exists at all energies in the massless theory $m=0$. However, we
should point out that even though in this case the gaugino
counterterm is present at very low energies, this does not mean
that the photino is massive. The reason for that is that in
massless QED the gauge coupling is renormalized to zero in the IR
and the physical photino mass is proportional to the fine
structure constant $\alpha$ which vanishes at low energies. This
is consistent with the fact that massless SQED should have
degenerate photons and photinos\foot{We conclude that in AdS the
photino remains massless both for nonzero and for zero $m$.  This
is consistent with general facts about AdS backgrounds. Massless
gauge fields are associated with conserved currents in the
boundary theory whose dimensions are fixed.  The dimension of the
superpartner of this current is therefore fixed by supersymmetry
and cannot be renormalized.  This dimension determines the mass of
the gauginos and hence they must remain massless.}.

It is easy to generalize this discussion in four different
directions:
 \item{1.} Consider SQED with several charged matter fields,
 flavors, with different masses. The coefficient of the counter
 term is proportional to the number of flavors which are lighter
 than the cutoff.  More precisely, it is proportional to the beta
 function at that energy.
 \item{2.} A non-Abelian gauge theory with matter fields also
 generates a gaugino counterterm which is proportional to its
 beta function at the scale of the UV cutoff.  Here, unlike SQED,
 loops of gauge multiplets also contribute to the counterterm.
 Therefore, at one loop order we can identify the coefficient of
 the counterterm \uonecontact\ as proportional to the one loop
 beta function coefficient $b_0$.
 \item{3.} The calculation above which leads to a term
 proportional to the constant $W_0$ is easily extended to
 include the full $W$.  Therefore, even if $W_0$ vanishes, the
 counterterm exists and leads to nontrivial interactions
 involving gauginos and scalars.
 \item{4.}  The previous discussion still holds when supersymmetry
 is spontaneously broken in flat space.  In this case we also find
 gaugino masses of order $\alpha \m3$ with a coefficient which
 depends on the beta function. However, in this case the
 renormalization group evolution of this term is different.  We
 can continue to integrate out modes to lower energies until we
 reach the gaugino mass and then this term stops running.
 Therefore, in that case the gauginos do receive nonzero physical
 mass.  This is the case of interest for hidden sector anomaly
 mediation models.

The phenomenon that a local, gauge non-invariant term is generated
by high momentum loops, is familiar in quantum field theory. For
example, in the massless Schwinger model, the vacuum polarization
is transverse:
 \eqn\pimunu{\Pi_{\mu \nu} \sim g_{\mu \nu} -{q_\mu q_\nu \over
 q^2}. }
Part of the local, $g_{\mu \nu}$ term arises from high energy
effects near the UV cutoff; the non-local term (and the remaining
contribution proportional to $g_{\mu \nu}$) is associated with
massless states in the loop. This as well as a similar phenomenon
in four dimensional four photon scattering are reviewed in
Appendix A.

Bagger et.al.\ \baggeretal\ have exhibited a similar structure in
supergravity theories which gives rise to the gaugino counterterm:
 \eqn\baggeroperator{\Gamma =
 -{g^2 \over 256 \pi^2} \int d^2 \theta 2 {\cal E} W^\alpha
 W_\alpha {1 \over \square}\left ( {\cal D}^2 - 9 R \right )
 \times \left [ 4 (T_R -3 T_G) R^+ - \dots \right ].}
This includes the local term
 \eqn\counterterm{{\cal L}_{ct} \sim {g^2 \over 16 \pi^2}W^*
 \lambda \lambda.}
The parallel to equation \pimunu\ is clear.

\newsec{A Transparent Derivation of the Counterterm}

The traditional derivation of the contact term, which we review
briefly below, relies on technical aspects of supergravity theory,
especially the use of the superconformal compensator.  Our
derivation of the previous section may seem tied to a particular
regularization scheme, and further arguments are required to
demonstrate that no additional local counterterms are possible. In
this section, we present a very simple derivation which relies
only on familiar field theory notions.

We start by reviewing the traditional derivation. The conformal
compensator is a non-propagating field introduced in a
supergravity formalism which begins with a conformally invariant
structure, and introduces a spurion (the compensator) to maintain
the symmetry. Following \refs{\murayamaetal,\randallsundrum}, we
write the compensator as:
 \eqn\compensator{\Phi = 1 + F_\Phi \theta^2}
In terms of $\Phi$, the relevant terms in the action are:
 \eqn\phiaction{\int d^4 \theta h(\bar \phi, e^{-V} \phi)
 \bar \Phi\Phi + \int d^2 \theta (\Phi^3 W(\phi) + f(\phi)
 W_\alpha^2) + {\rm h.c.}}
$h$ is related to the K\"ahler potential through
 \eqn\fkahler{h= -3 M_p^2 e^{-K/3M_p^2}.}
For a model with broken supersymmetry and vanishing cosmological
constant:
 \eqn\fphiequals{F_\Phi = m_{3/2}.}
In our simple $U(1)$ theory, if we regulate with a Pauli-Villars
field with mass $\Lambda$, the gauge coupling function at the
scale $\mu$ is:
 \eqn\fpvhiggs{-\int d^2 \theta {b_0 \over 32 \pi^2}
 \ln\left({\mu^2 \over \Lambda^2 \Phi^{-2}}\right)W_\alpha^2.}
where the power of $\Phi$ in the logarithm compensates for the
dimension of $\Lambda$. Expanding the logarithm and doing the
$\theta$ integration, generates precisely the gaugino contact
term.

But there is a conceptually much simpler derivation, not tied to
any particular supergravity formalism, or to anomalies in field
redefinitions.  Consider, again, the $U(1)$ theory, with massless
chiral fields.  This theory has a Higgs phase in which:
 \eqn\phivev{\phi_+ = \phi_- = v}
(up to a phase).  In this phase, the gauge symmetry is broken. The
vector multiplet is massive, and there is one massless chiral
multiplet, which can be described by the gauge invariant composite
field $\phi_+\phi_-$. In perturbation theory, there are no
couplings of two light fields to heavy fields, so there is no
interesting infrared behavior in Feynman diagrams. The effective
action for the gauge fields is necessarily local, so it has the
form:
 \eqn\waction{\int d^2 \theta  \left({1 \over g^2} +
 {b_0 \over 32 \pi^2} \ln(\phi_+\phi_-/\Lambda^2)
 \right)W_\alpha^2.}
(Of course, this term can be understood as reflecting the anomaly
in the global symmetry under which $\phi_\pm$ are rotated by the
same phase, or the conformal anomaly of SQED.)  This corresponds
to a gauge coupling function,
 \eqn\gaugefunction{f= {b_0 \over 32 \pi^2}\ln(\phi_+ \phi_-)+
 const.}
Substituting this into the general supergravity action
\refs{\cremmer,\wessbagger}, this gauge coupling function leads to
a gaugino contact term:
 \eqn\higgscontactterm{{1 \over M_p^2}
 \lambda \lambda g^{i \bar i} {\partial f \over \partial
 \phi_i} ({D_i W})^*= {1 \over 16 \pi^2M_p^2} \lambda \lambda
 W_0^*+ ....}
where $D_i W$ denotes the K\"ahler derivative of the
superpotential with respect to the chiral field $\phi_i$:
 \eqn\dw{D_i W = {\partial W \over \partial \phi_i} + {1 \over M_p^2}
 {\partial K \over \partial \phi_i} W.}
Unlike the discussion in the previous section, here the gaugino
term is part of a supersymmetric Lagrangian.

This Higgs phase analysis makes it absolutely clear that the term
is necessary for supersymmetry, and allows one to immediately
write the complete action.

We see that the gaugino counterterm can be supersymmetrized either
using a nonlocal action as in \baggeretal, or using a singular
action like \gaugefunction, but it cannot be supersymmetrized
using a local regular action.  In the previous section we worked
around the origin in field space $\phi_\pm \approx 0$, and
therefore we could not use expressions like \gaugefunction.  Here,
in the Higgs phase, we are far from the origin, and therefore we
can use this expression.  More physically, the subtleties in the
previous section are associated with massless particles (more
precisely, particles which are much lighter than the UV cutoff).
In the Higgs phase, there are no such particles and therefore
these subtleties are absent and the term can be supersymmetrized
by local operators.

The Higgs phase calculation indicates most strikingly that the
gaugino contact term is not related to an anomaly.  In a theory
without charged massless fields in its Higgs phase, it is {\it
required} by supersymmetry.  We will see shortly that similar
remarks apply to scalar mass terms.  First, we demonstrate that
the contact term arises already in globally supersymmetric
theories.

\newsec{The Counterterm in Globally Supersymmetric Theories}

With the machinery of the previous section, we can see that the
gaugino counterterm already arises in globally supersymmetric
theories. Take the $U(1)$ model as before, but include also a
Kahler potential:
 \eqn\singlefieldk{K=\bar  \phi_+ \phi_+ + \bar  \phi_- \phi_- +
 \bar z z + {1 \over \mu^2} (\bar\phi_+ \phi_++
 \bar\phi_- \phi_-) \bar z z.}
Here $z$ is a field with a non-zero F component whose dynamics is
not important here (see section 5), and $\mu$ is some energy scale
assumed far lower than $M_p$, so gravity is irrelevant The last
term in \singlefieldk\ can arise in a more microscopic
renormalizable theory from tree level exchange of massive gauge
fields, or from loop effects. Again, consider the theory in its
Higgs phase. The one loop effective action has the structure:
 \eqn\oneloophiggs{{\cal L}_0 + {b_0 \over 16 \pi^2} \int d^2
 \theta \ln (\phi_+ \phi_-) W_\alpha^2.}
Solving for $F_{\phi_+}$ and $F_{\phi_-}$ gives
 \eqn\fphis{F_{\phi_+} = -{1 \over \mu^2} \phi_+ \bar z F_z \qquad ;
 \qquad F_{\phi_-} = -{1 \over \mu^2} \phi_- \bar z F_z.}
So again, substituting in  \oneloophiggs\ yields a gaugino mass
contact term:
 \eqn\globalsusy{{2 b_0\over 16 \pi^2 \mu^2} \lambda \lambda \bar z
 F_z.}

This term has all of the features of the counterterm in
supergravity theories. In the theory with $\phi_\pm=0$, it cannot
be written as part of a locally supersymmetric effective action.
Its appearance is required by supersymmetry, but how it appears
depends on the choice of regulator.  For example, with a
Pauli-Villars regulator, it may be calculated directly, but it is
not generated with a momentum space cutoff or by dimensional
reduction, and so must be added by hand in these cases.

An alternative derivation of the answer \globalsusy\ which is
valid around the origin $\phi_\pm \approx 0$ can be obtained by
performing a field redefinition $\phi_\pm \to \phi_\pm (1 + {\bar
z z \over 2 \mu^2})$.  This rescaling is not holomorphic but this
is not a problem.  The anomaly in this rescaling leads to a term
proportional to $\int d^2 \theta \log(1 + {\bar z z \over \mu^2})
W_\alpha^2 \supset {\bar z F_z\over \mu^2 }\lambda \lambda$.

This derivation is easily generalized to an arbitrary K\"ahler
potential which depends on fields in different representations of
the gauge group. Then it leads to a term proportional to
 \eqn\geneK{ \partial_j \left(\sum_{R} {T_R \over d_R} \log
 (\det_{i \bar i} K_{i \bar i}) \right) F^j \lambda\lambda,}
where the sum over $R$ is over the different representations,
$T_R$ and $d_R$ are the Casimir and dimension of $R$, and the
indices $i $ and $\bar i$ label fields in $R$ and $\bar R$.  The
term \geneK\ has already been noted in \baggeretal.  However,
these authors have set $M_p=1$, and therefore did not stress that
this term is independent of $M_p$ and hence it is unrelated to
gravity!

\newsec{A setting:  Supergravity Theories with Dynamical
Supersymmetry Breaking}

With a view to thinking about scalar contact terms, in this
section we study a simple model for supersymmetry breaking.  Many
(but not all) models of tree level or dynamical supersymmetry
breaking are described at low energies by this model or a simple
variant of it.  We start by considering a global supersymmetric
theory and later we will couple it to supergravity and use it as a
hidden sector for supersymmetry breaking.

We have a single chiral superfield $z$ with a K\"ahler potential
and a superpotential
 \eqn\KWftg{\eqalign{
 &K_{hidden}=  \mu^2 f\left({z\bar z\over  \mu^2}\right)\cr
 &W_{hidden} = M^2 z+ W_0.}}
for some function $f(z \bar z/ \mu^2) $.  In models of dynamical
supersymmetry breaking the scale $\mu$ is the dynamically
generated scale of the theory, and it determines the low energy
K\"ahler potential. In order for this effective theory to be
valid, we need that the scale of supersymmetry breaking is much
smaller than $\mu$ and hence we take $M \ll \mu$. Therefore, we
took the K\"ahler potential to be independent of $M$.  Note that
in global supersymmetry the constant $W_0$ is not important and
hence the theory has a $U(1)_R$ symmetry under which $z$ rotates
by a phase. Such a symmetry is common in models of supersymmetry
breaking \refs{\AffleckXZ,\NelsonNF}.

The potential derived from \KWftg\ is
 \eqn\Vft{V_{hidden} ={M^4 \over
 f''( z\bar z/\mu^2) {z\bar z \over \mu^2} + f' ( z\bar
 z/\mu^2)}.}
If the function $f$ is regular, the potential never vanishes and
supersymmetry is broken (the behavior of $f$ at infinity
determines whether or not the theory has runaway behavior).  It
leads to $F_z \sim M^2$.  If the minimum of the potential is at
nonzero $z$ (which is necessarily at $z \sim \mu$), then the
$U(1)_R$ symmetry is spontaneously broken.\foot{The $(3,2)$ model
of \AffleckXZ\ and its various relatives lead to a similar but
somewhat more complicated situation.  There as the analog of $M$
is reduced, the expectation values of the low energy fields become
larger rather than remaining constant as in our model.}
Alternatively, as in the O'Raifeartaigh model and in the model of
\IntriligatorDD\ the minimum can be at $z=0$, and then the
R-symmetry is not broken. In this case it is enough to expand $f$
and study
 \eqn\Khs{K_{hidden}= z \bar z -  {z^2 \bar z^2 \over \mu^2} }
which leads to $F_z=M^2$ and the field $z$ acquires a mass  $m_z
\sim M^2/\mu \ll M\ll \mu$.

Next we consider the gravitational corrections to these
expressions focusing on the weak gravity limit $M_p \to \infty$.
For the above field theoretic analysis to be meaningful we take
$M,\mu \ll M_p$.  Rather than taking $M_p\to \infty$ with fixed
$M,\mu$, we consider the limit with fixed gravitino mass
 \eqn\gravimass{\m3 \sim {M^2 \over M_p} ,}
and therefore we take
 \eqn\Mmulim{M,\mu\sim \sqrt{M_p}\to \infty.}
The supergravity potential
 \eqn\Vgen{V=e^{K/M_p^2}\left({1\over K_{z\bar z}} \big|\partial_z
 W + {1\over M_p^2} K_z W\big|^2 - {3 \over
 M_p^2}\big|W\big|^2\right)}
can be analyzed in the limit $M_p \to \infty$ with \Mmulim. In
order to cancel the orders $M_p^2$ and $M_p$ contributions to the
cosmological constant,
 \eqn\Wco{W_{hidden} =
 M^2 \left( z + {1 \over \sqrt 3}M_p - { \mu^2 \over
 6 \sqrt 3 M_p} + \CO(1/M_p)  \right)}
(recall \Mmulim). Order $M_p^0$ effects in the cosmological
constant and in $W_0$ depend on the higher order corrections to
$K$ and on quantum effects in the low energy visible theory. We
will not discuss them here.

Note that the constant term in $W$ explicitly breaks the $U(1)_R$
symmetry. Since this constant is $\CO(M_p^2)$ one should check
that the previous results about the field $z$ are not modified.
Indeed, to the order we work the only difference due to the change
in $W$ and the gravitational corrections is to change the
expectation value of $z$ to
 \eqn\zvev{\langle z \rangle = z_0= {\mu^2 \over 2\sqrt 3 M_p}
 (1 +\CO(1/M_p))}
but leaving the leading order result for the mass of $z$ as in the
field theory calculation.  Note that in our limit the expectation
value \zvev\ is of order $M_p^0$, but since the gravitational
corrections are of order $1 \over M_p^2$ our analysis is
consistent.

Finally, the gravitino mass is given by
 \eqn\gravitinomass{\m3={1 \over M_p^2} e^{K\over 2M_p^2} W(z_0)
 ={M^2 \over \sqrt 3 M_p} (1 + \CO(1/M_p))}
in accord with \gravimass.

\newsec{Scalar Field Contact Terms}

One of the remarkable observations of \randallsundrum\ is that not
only are there contact terms for gauginos which seem incompatible
with local supersymmetry, but there are also scalar terms.  Such
scalar mass terms are conceptually similar to the gaugino masses
we discussed above.  The lesson from our previous analysis is that
despite appearance, there is really no tension between these
masses and the local supersymmetry.

The most transparent way to understand it is as in our analysis of
the gaugino contact term in the Higgs phase of the theory.  In
that case, there is no issue of non-locality and it is clear that
supersymmetry demands the presence of the contact term. The same
is true of the scalar masses, as we will see in this section.

The gaugino counterterm is most useful in hidden sector
supergravity theories without gauge singlets.  In this case the
tree level couplings lead to very small gaugino masses and the
gaugino counterterm is the leading contribution. Scalar masses, on
the other hand, are easy to generate in all hidden sector theories
using generic dimension four operators in the K\"ahler potential
which couple the hidden and the visible sectors.  In such a situation,
the masses which are generated by the local counterterms are
suppressed by powers of the fine structure constant, and hence
they are negligible. Randall and Sundrum \randallsundrum\
considered a certain ``sequestered" form of the K\"ahler
potential, which might arise in some contexts (particularly in the
case of separated branes).  This form guarantees that the tree
level terms (and also the one loop counterterms) do not lead to
scalar masses.  The leading contribution to the scalar masses
arises from a two-loop counterterm.  In the rest of this section
we will describe this phenomenon using our Higgs phase langauge.

We divide the fields into two groups, visible sector fields $\phi$
($\phi_+$ and $\phi_-$ in our $U(1)$ example), and hidden sector
fields, $z$.  For simplicity we consider a single hidden sector
field, as in section 5, and a single visible sector field $\phi$.
Instead of the interaction with the $U(1)$ gauge field we can have
$\phi^3$ interaction in the visible sector superpotential.  The
extension to a $U(1)$ theory is straightforward.

The sequestered K\"ahler potential is
 \eqn\sequestered{
 K= -3 M_p^2\ln\left (1-{1\over 3 M_p^2} K_{vis}(\phi,\bar \phi)
  - {1\over 3 M_p^2} K_{hid}(z, \bar z)\right).}
For the hidden sector we use the model of section 5 and tune $W_0$
to have vanishing cosmological constant.

More concretely, the visible and the hidden sectors K\"ahler
potentials $K_{vis}$ and $ K_{hid}$ in \sequestered\ are such that
 \eqn\specifick{
 K = -3 M_p^2 \ln\left(1-{1 \over 3M_p^2}\bar\phi \phi Z(\bar\phi
 \phi) - {1 \over 3 M_p^2} (\bar z z - {1 \over \mu^2}
 z^2 \bar z^2)\right)}
and the superpotential is
 \eqn\superpotential{ W = W_0 + M^2 z + W_{vis}(\phi).}
At tree level $Z(\phi,\bar\phi)=1$.  Radiative corrections in the
visible sector change $Z$, but the important point is that it is
independent of gravitational corrections.

At tree level $Z=1$ and the scalar fields $\phi$ do not get
supersymmetry breaking mass terms \randallsundrum.  One can then
do a calculation of the corrections to the masses, using
Pauli-Villars regulators as for the gaugino masses. The
Pauli-Villars fields have a non-zero supersymmetry-breaking mass
(B-term).  Since $\mu \ll M_p$, this is simply:
 \eqn\bterm{\Lambda m_{3/2} \phi^2 + {\rm c.c.}.}
So the mass matrix for these fields is precisely that of a
gauge-mediated theory (for a review, see e.g.\ \gaugemediation),
and we can immediately read off the two loop correction to the
masses of the light fields:
 \eqn\twoloopmasses{m_s^2 = -2 \left ({\alpha \over 16 \pi^2}
 \right )^2}
Here we have given the expression when the visible sector is SQED which
has two chiral superfields $\phi_\pm$.  In the case of a single
$\phi$ with a $\phi^3$ superpotential interaction $\alpha$ in
\twoloopmasses\ is replaced by the square of the cubic coupling.
As for the gaugino mass, the result is independent of the
Pauli-Villars mass $\Lambda$.  As there, it cancels the
corresponding contributions from physical heavy fields, and we are
left with the Paul-Villars contribution associated with the light
fields. The negative sign comes from the need to {\it subtract}
the contribution of the Pauli-Villars fields.

All of this is precisely analogous to the behavior we saw for the
gaugino mass.  In the conformal compensator approach, these masses
arise, as in that case, from thinking of the ultraviolet cutoff as
dependent on the compensator \randallsundrum. Once more, these
results can be understood in terms of the appearance in the
Wilsonian effective action of a counterterm which does not respect
the local supersymmetry.  The lack of local SUSY invariance is
needed in order to compensate the lack of invariance of the
measure of the light fields.  The required supergraph calculation
in this case is more challenging (one needs the terms in the
lagrangian quadratic in the auxiliary fields in the gravity
multiplet, for example).

A Higgs phase calculation similar to the one we used for the
gaugino masses is only slightly more complicated than in that
case. As we now illustrate, the scalar masses follow from
straightforwardly computing the K\"ahler potential, and then using
the supergravity action to determine the scalar potential.

In the Higgs phase, $Z$ receives $\phi$-dependent radiative
corrections. Including the leading logarithms up to two loops we
have
 \eqn\zequals{Z= 1+ a_1 \epsilon \ln(\bar\phi
 \phi) + a_2 \epsilon^2 \ln^2(\bar\phi \phi).}
Here $\epsilon = {g^2 \over 16 \pi^2}$ or the square of the cubic
coupling in the superpotential.   The coefficients $a_1$ and $a_2$
can be read off of standard calculations in supersymmetry (see,
e.g., \martinvaughn).  In the case of SQED we have:
 \eqn\epsilondelta{a_1 =
 1\qquad ; \qquad a_2 = 1-2 b_0}
(that case needs several charged fields like $\phi_\pm$) and other
values for the Wess-Zumino model with the cubic superpotential.

Next, we substitute this expression for $Z$ in the K\"ahler
potential \specifick\ and then in the supergravity scalar
potential.  Using the expectation value of $ z$ from \zvev\ and
tuning $W_0$ so that the cosmological constant vanishes we
determine the potential for $\phi$
 \eqn\Vgen{\eqalign{
 V(z_0)= & \ e^{K/M_p^2}\left({g^{i\bar i}} \big|\partial_i
 W + {1\over M_p^2} \partial_i K W\big|^2 - {3 \over
 M_p^2}\big|W\big|^2\right)\cr
 = &{1\over \partial_\phi\partial_{\bar\phi}
 K_{vis}(\phi,\bar\phi)} |\partial_\phi W_{vis}(\phi)|^2\cr
 &+ \m3^2 \epsilon^2(a_1^2 - 2a_2 )|\phi|^2\cr
 & + \m3 \left( \Delta_\phi \phi
 \partial_\phi W_{vis}(\phi) - 3 W_{vis}(\phi) + c.c.
  \right)\cr
 & + \CO(1/M_p, \epsilon^3)\cr
 \Delta_\phi= & 1- a_1 \epsilon + \CO(\epsilon^2)
 }}
(recall, $\mu,M \sim \sqrt {M_p}$.)

The first term in \Vgen\ is the potential for $\phi$ in the
globally supersymmetric limit. The corrections represent
supersymmetry breaking terms.

Consider first the scalar masses of the form $|\phi|^2$ (the
second term in \Vgen). As claimed in \randallsundrum, the
sequestered form does not lead to tree level masses of order
$\epsilon^0$. Also, the one loop correction of order $\epsilon^1$
and the two loop contributions which could depend on logarithms
like $ \epsilon^2 \ln^2(\bar\phi \phi)$ and $\epsilon^2
\ln(\bar\phi \phi)$ vanish. We are left with a two loop mass term
without logarithms.  Such an answer can be extrapolated to $\phi
\approx 0$ where it leads to the scalar mass square $ m_s^2=(a_1^2
- 2 a_2 ) \epsilon^2 \m3^2 $. This agrees with the expression of
Randall and Sundrum for the scalar masses:
 \eqn\scalarmasses{m_s^2 =2 b_0\left ({g^2
 \over 16 \pi^2} \right )^2m_{3/2}^2.}

The third term in \Vgen\ leads to B-terms and A-terms.  The
B-terms arise already at tree level (as in section 2). But the
A-terms which originate from $\phi^3$ terms in the superpotential
arise at one loop and are of order $\epsilon$. We expressed them
in terms of the anomalous dimension $\Delta_\phi$. \foot{The
$\CO(\epsilon^2)$ terms in $\Delta_\phi$ depend on $\log
|\phi|^2$.  These terms should be understood after performing
wavefunction renormalization.} This way of writing them can be
used to make contact with the formalism based on the conformal
compensator.

Once again, in this formulation, the scale dependence of the mass,
A-terms and B-terms is immediate.  It is also clear, once more,
that these terms are required by supersymmetry.

\appendix{A}{Two Familiar Analogs of the Gaugino Counter Term}

\subsec{Contact Terms in the Schwinger Model}

Electrodynamics in $1+1$ dimension poses many of the same issues
which arise with the gaugino counterterm. A traditional way of
describing mass generation in the Schwinger model is to examine
the vacuum polarization diagram.  The vacuum polarization itself
is finite, but the diagram is superficially ultraviolet divergent,
and this can lead to paradoxes.  For example, it is easy to
``prove" that the vacuum polarization tensor vanishes.  Writing
the transverse expression
 \eqn\transversality{ \Pi_{\mu \nu}(q) = (g_{\mu \nu} q^2 - q_\mu
 q_\nu) \Pi(q^2)}
one can take the trace:
 \eqn\pitrace{\Pi^\mu_\mu = q^2 \Pi(q^2).}
But at the level of Feynman diagrams for massless fields, this
would seem to vanish since $\gamma^\mu \gamma^\nu \gamma_\mu =0$
in two dimensions. However, the diagram is ultraviolet divergent
by power counting, and introduction of a gauge-invariant regulator
resolves the puzzle.  For example, for dimensional regularization,
$\gamma^\mu \gamma^\nu \gamma_\mu = \epsilon$ and, combined with
the $1/\epsilon$ from the ultraviolet divergence, yields a finite
contribution.  Alternatively, with a Pauli-Villars regulator, one
obtains two contributions. {}From the diagram with the massless
fields, one obtains:
 \eqn\masslesscontribution{ \Pi^0_{\mu \nu}(q)
 =  (2q_\mu q_\nu -g_{\mu \nu} q^2){1 \over \pi q^2},}
while from the regulator diagram one obtains
 \eqn\regulatorcontribution{ \Pi^\Lambda_{\mu \nu}(q) = -g_{\mu
 \nu} q^2{1 \over \pi q^2}.}
(In lightcone coordinates $ \Pi^0_{\pm\pm}(q) =  {q_\pm  \over \pi
q_\mp} $, $\Pi^0_{\pm\mp}=0$, while  $ \Pi^\Lambda_{\pm\pm}(q)=  0
$, $\Pi^\Lambda_{\pm\mp}=-{1\over \pi}$). Note that neither result
by itself is gauge invariant, but the combined expression is.
Taking account of the normalization of the kinetic terms, this
corresponds to a mass, $e^2/\pi$, for the physical excitation.

A few comments are in order.  First, as for the gaugino
counterterm, there is a local piece in this expression, arising
from high energy modes, and there is a non-local term, from
massless exchanges, which compensates for the lack of gauge
invariance of the contact term.  Second, it is important to point
out that a failure of gauge invariance (breakdown of the Ward
identity), can be understood, from a path integral perspective, as
resulting from a lack of invariance of the measure.  The naive
measure, without the regulator field, violates gauge invariance.
The regulated measure does not. This violation of gauge invariance
{\it has nothing to do with whether the fields are massless or
massive}.  In the massive theory without the regulator, for a
fermion of mass $m$, we would obtain a result identical to that
above for $\Pi^\Lambda$.

Had one used a non-gauge invariant regulator, such as a momentum
space cutoff, one would need to fix up the short distance part by
adding a counterterm (i.e.\ a piece of the high energy Wilsonian
action) to the contribution from the Feynman diagram.

\subsec{A four dimensional example:  light by light scattering}

Consider now four dimensional, vector-like electrodynamics (with
massless fermions). Here, there is no divergence associated with
diagrams with four external photons. This is because of gauge
invariance. But it is not true that the high energy behavior of
these diagrams can be ignored.

Analogous to the Schwinger model, one expects that the 1PI action
at low energies contains terms like
 \eqn\lfourphoton{ {\cal L} =
 {(F_{\mu \nu}^2)(F_{\rho \sigma}^2) \over \square^2}.}
In momentum space, this includes couplings like
 \eqn\equationa{ A_\mu^2 A_\nu^2\qquad ;\qquad {q_\mu A^\mu
 q_\nu A^\nu A_\rho^2 \over q^2} }
and so on.

It is easy to see the role of short distances in the Feynman
graphs. In the one loop graph, with four external gauge bosons,
with polarization indices $a,b,c,d$, the graph behaves in the
ultraviolet as:
 \eqn\manyslashes{ \int  {d^4 p \over (2 \pi)^4}
 {{\rm Tr} (\gamma^a \pslash \gamma^b \pslash  \gamma^c \pslash
 \gamma^d \pslash + \gamma^a \pslash \gamma^b \pslash  \gamma^d
 \pslash \gamma^c \pslash + \gamma^a \pslash \gamma^c \pslash
 \gamma^c \pslash \gamma^d \pslash) \over (p^2)^4 }. }
We can make the further simplification of contracting with $g_{ab}
g_{cd}$. Then the integrand vanishes. However, it is necessary to
introduce a regulator. In dimensional regularization, for example,
the result is:
  \eqn\dimregresult{
  {1 \over (2\pi)^d} \int  d^d p~ {1 \over p^4}~ {2 \epsilon}
  = {2 \over 16 \pi^2}.}
So there is a finite contact term of the type suggested above. The
calculation is precisely analogous to that of the Schwinger model.

In the case of a massive field, the situation is parallel to that
of the Schwinger model.  Before regularization, there appears to
be a local contact term. The integrand is now proportional to:
 \eqn\nonlocalct{\eqalign{
  \int  d^4 p\, {\rm Tr} \Big[ \gamma^a (\pslash + m)
 \gamma^b (\pslash + m) & \gamma^c (\pslash +m )\gamma^d (\pslash +
 m) \gamma^a (\pslash + m) \gamma^b (\pslash + m)  \gamma^d
 (\pslash + m) \gamma^c (\pslash + m) \cr
  &+  \gamma^a (\pslash + m) \gamma^c (\pslash + m)  \gamma^c
  \pslash \gamma^d \pslash \Big] / (p^2-m^2)^4 .
 }}
Contracting with $g_{ab}$ and $g_{cd}$ as before, and working
directly in four dimensions, the trace in the numerator becomes
$24 m^4$ (again, there is no divergence). So one is left with a
finite, local interaction term:
 \eqn\fourgammacontact{ {\cal L}_{4A} = {1 \over 16 \pi^2}
 A_\mu A^\mu A_\nu A^\nu.}
This term is not gauge invariant.  Introducing a regulator cancels
the contact term, leading to a gauge invariant result (the famous
Euler-Heisenberg lagangian).  In the case of a massless field,
there is no such cancellation, but now there is a non-local term
in the 1PI effective action, whose gauge-non-invariance cancels
that of the contact term, just as in the supergravity case.

\bigskip
 \centerline{\bf Acknowledgements}
We thank Nima Arkani-Hamed, Tom Banks, Linda Carpenter, Juan
Maldacena, Scott Thomas and Edward Witten for conversations. The
research of NS was supported in part by DOE grant
DE-FG02-90ER40542. M.D. enjoyed the hospitality of the Kavli
Institute for Theoretical Physics while this work was under way.

\listrefs
\end